\documentstyle[aps,psfig,preprint,tighten,floats]{revtex}
\newcommand{\gsim}{\lower2pt\hbox{$\:\stackrel{>}{\scriptstyle\sim}\:$}}
\newcommand{\lsim}{\lower2pt\hbox{$\:\stackrel{<}{\scriptstyle\sim}\:$}}
\begin{document}
\draft
\title{Dynamics of localization in a waveguide}
\author{C.W.J. Beenakker}
\address{
Instituut-Lorentz, Universiteit Leiden\\
P.O. Box 9506, 2300 RA Leiden, The Netherlands}
\date{September 2000}
\maketitle
\begin{abstract}
This is a review of the dynamics of wave propagation through a disordered
$N$-mode waveguide in the localized regime. The basic quantities considered are
the Wigner-Smith and single-mode delay times, plus the time-dependent power
spectrum of a reflected pulse. The long-time dynamics is dominated by resonant
transmission over length scales much larger than the localization length. The
corresponding distribution of the Wigner-Smith delay times is the Laguerre
ensemble of random-matrix theory. In the power spectrum the resonances show up
as a $t^{-2}$ tail after $N^{2}$ scattering times. In the distribution of
single-mode delay times the resonances introduce a dynamic coherent
backscattering effect, that provides a way to distinguish localization from
absorption.
\medskip\\
{\sf To appear in: Photonic Crystals and Light Localization,\\
edited by C.M. Soukoulis,
NATO Science Series (Kluwer, Dordrecht, 2001).}
\end{abstract}
\narrowtext

\section{Introduction}
\label{intro}

Light localization, one of the two central themes of this meeting, has its
roots in electron localization. Much of the theory was developed first for
electrical conduction in metals at low temperatures, and then adapted to
propagation of electromagnetic radiation through disordered dielectric media
\cite{She90,Tig99}. Low-temperature conduction translates into propagation that
is monochromatic in the frequency domain, hence static in the time domain.

This historical reason may explain in part why much of the literature on
localization of light deals exclusively with static properties. Of course one
can think of other reasons, such as that a laser is a highly monochromatic
light source. It is not accidental that one of the earliest papers on wave
localization in the time domain \cite{Whi87} appeared in the context of
seismology, where the natural wave source (an earthquake or explosion) is more
appropriately described by a delta function in time than a delta function in
frequency.

Our own interest in the dynamics of localization came from its potential as a
diagnostic tool. The signature of static localization, an exponential decay of
the transmitted intensity with distance, is not unique, since absorption gives
an exponential decay as well \cite{Lag00}. This is at the origin of the
difficulties surrounding an unambiguous demonstration of three-dimensional
localization of light \cite{Nat99}. The dynamics of localization and absorption
are, however, entirely different. One such dynamical signature of localization
\cite{Sch00} is reviewed in this lecture.

\begin{figure}[tb]
\centerline{
\psfig{figure=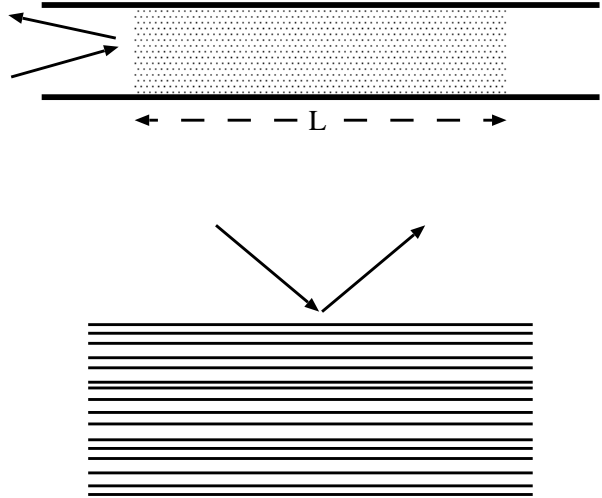,width= 8cm}}
\medskip
\caption[]{
The top diagram shows the quasi-one-dimensional geometry considered in this
review. The waveguide contains a region of length $L$ (dotted) with randomly
located scatterers that reflects a wave incident from one end (arrows). The
number of propagating modes $N$ may be arbitrarily large. The one-dimensional
case $N=1$ is equivalent to the layered geometry shown in the bottom diagram.
Each of the parallel layers is homogeneous but differs from the others by a
random variation in composition and/or thickness.
\label{wavevslayer}
}
\end{figure}

Localization is a non-perturbative phenomenon and this severely complicates the
theoretical problem. In two- and three-dimensional geometries (thin films or
bulk materials) not even the static case has been solved completely
\cite{Efe97}. The situation is more favorable in a one-dimensional waveguide
geometry, where a complete solution of static localization exists
\cite{Efe97,Bee97}. The introduction of dynamical aspects into the problem is a
further complication, and we will therefore restrict ourselves to the waveguide
geometry (see Fig.\ 1). The number $N$ of propagating modes in the waveguide
may be arbitrarily large, so that the geometry is more appropriately called
{\em quasi}-one-dimensional. (The strictly one-dimensional case $N=1$ is
equivalent to a layered material.)

The basic dynamical quantity that we will consider is the auto-correlator of
the time-dependent wave amplitude $u(t)$,
\begin{equation}
a_{\omega}(t)=\int_{-\infty}^{\infty}dt'\,e^{-i\omega
t'}u(t)u(t+t').\label{Comegatdef}
\end{equation}
If the incident wave is a pulse in time, then the transmitted or reflected wave
consists of rapid fluctuations with a slowly varying envelope (see Fig.\ 2).
The correlator $a_{\omega}(t)$ selects the frequency component $\omega$ of the
rapid fluctuations. The remaining $t$-dependence is governed by the propagation
time through the waveguide.

\begin{figure}[tb]
\centerline{
\psfig{figure=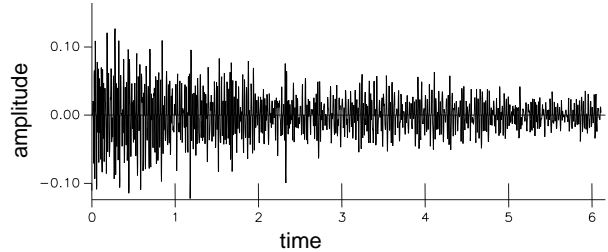,width=
8cm}}
\medskip
\caption[]{
Computer simulation of an acoustic plane wave pulse reflected by a randomly
layered medium. The medium is a model for the subsurface of the Earth, with a
sound velocity that depends only on the depth. The figure shows the reflected
wave amplitude as a function of time (arbitrary units). The incident pulse
strikes the surface at time zero. From Ref.\ \protect\cite{Asc91}.
\label{georgep}
}
\end{figure}

If the incident wave is not a pulse in time but a narrow band in frequency,
then it is more convenient to study the frequency correlator
\begin{equation}
a_{\omega}(\delta\omega)=\int_{-\infty}^{\infty} dt\,e^{i\delta\omega
t}a_{\omega}(t)=u^{\ast}(\omega)u(\omega+\delta\omega).
\label{Comegadeltaomegadef}
\end{equation}
The Fourier transformed wave amplitude $u(\omega)=\int dt\, e^{i\omega
t}u(t)\equiv I^{1/2}e^{i\phi}$ is complex, containing the real intensity
$I(\omega)$ and phase $\phi(\omega)$. Most of the dynamical information is
contained in the phase factor, which winds around the unit circle at a speed
$d\phi/d\omega$ determined by the propagation time (see Fig.\ 3).

\begin{figure}[tb]
\centerline{
\psfig{figure=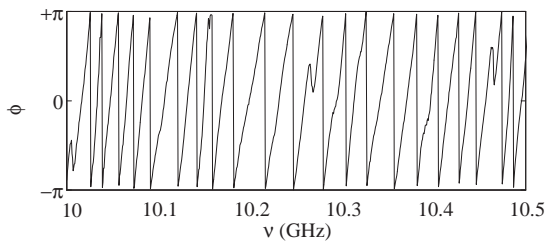,width=
8cm}}
\caption[]{
Frequency dependence of the phase (modulo $2\pi$) of microwave radiation
transmitted through a disordered waveguide. The waveguide consists of a 1~m
long, 7.6~cm diameter copper tube containing randomly positioned polystyrene
spheres (1.27~cm diameter, 0.52\% volume filling fraction). Wire antennas are
used as the emitter and detector at the two ends of the tube. From Ref.\
\protect\cite{Seb99}.
\label{sebbah}
}
\end{figure}

The correlator $a$ depends sensitively on the random locations of the
scatterers in the waveguide, that give rise to the localization. This calls for
a statistical treatment, in which we consider the probability distribution of
$a$ in an ensemble of waveguides with different scatterer configurations. The
method of random-matrix theory has proven to be very effective at obtaining
statistical distributions for static scattering properties \cite{Bee97}. The
extension to dynamical properties reviewed here is equally effective for
studies of the reflected wave. The time dependence of the transmitted wave is
more problematic, for reasons that we will discuss.

\section{Low-frequency dynamics}
\label{sec2}

The low-frequency regime is most relevant for optical and microwave experiments
\cite{Lag00,Seb99,Gen99}, where one usually works with an incident beam that
has a narrow frequency bandwidth relative to the inverse propagation time
through the system. We assume that the length $L$ of the waveguide is long
compared to the (static) localization length $\xi=Nl$, which is equal to the
product of the number of propagating modes $N$ and the mean free path $l$. The
reflected wave amplitudes $r_{mn}$ in mode $m$ (for unit incident wave
amplitude in mode $n$) are contained in an $N\times N$ reflection matrix $r$.
This matrix is unitary, provided we can disregard absorption in the waveguide.
It is also symmetric, because of reciprocity. (We do not consider the case that
time-reversal symmetry is broken by some magneto-optical effect.)

The correlator
\begin{equation}
C_{\omega}(\delta\omega)=r^{\dagger}(\omega)r(\omega+\delta\omega)
\label{Comegardef}
\end{equation}
is the product of two unitary matrices, so it is also unitary. Its eigenvalues
$\exp(i\phi_{n})$, $n=1,2,\ldots N$, contain the phase shifts $\phi_{n}$. Since
$\phi_{n}\equiv 0$ for all $n$ if $\delta\omega=0$, the relevant dynamical
quantity at low frequencies is the limit
\begin{equation}
\tau_{n}=\lim_{\delta\omega\rightarrow
0}\frac{\phi_{n}}{\delta\omega},\label{taundef}
\end{equation}
which has the dimension of a time. It is known as the Wigner-Smith delay time,
after the authors who first studied it in the context of nuclear scattering
\cite{Wig55,Smi60}. The $\tau_{n}$'s may equivalently be defined as the
eigenvalues of the Hermitian time-delay matrix $Q$,
\begin{equation}
Q(\omega)=-ir^{\dagger}\frac{dr}{d\omega}=U^{\dagger}{\rm
diag}(\tau_{1},\tau_{2},\ldots\tau_{N})U.\label{Qdef}
\end{equation}

Experiments typically measure not the product of matrices, as in Eq.\
(\ref{Comegardef}), but the product of amplitudes, as in Eq.\
(\ref{Comegadeltaomegadef}). The amplitude measured within a single speckle (or
coherence area) corresponds to a single matrix element. The typical observable
is therefore not the Wigner-Smith delay time but a different dynamical quantity
called the single-channel (or single-mode) delay time \cite{Seb99,Gen99}:
\begin{equation}
\tau_{mn}=\lim_{\delta\omega\rightarrow 0}{\rm
Im}\frac{r_{mn}^{\ast}(\omega)r_{mn}^{\vphantom\ast}(\omega+\delta\omega)}
{\delta\omega|r_{mn}(\omega)|^{2}}.\label{taumndef}
\end{equation}
If we decompose the complex reflection amplitude into intensity and phase,
$r_{mn}=I^{1/2}e^{i\phi}$, then the single-channel delay time is the phase
derivative, $\tau_{mn}=d\phi/d\omega\equiv\phi'$. Since the reflection matrix
$r(\omega+\delta\omega)$ has for small $\delta\omega$ the expansion
\begin{equation}
r_{mn}(\omega+\delta\omega)=\sum_{k}U_{km}U_{kn}(1+i\tau_{k}\delta\omega),
\label{rexpansion}
\end{equation}
we can write the single-channel delay time as a linear combination of the
Wigner-Smith times,
\begin{equation}
\tau_{mn}\equiv\phi'={\rm
Re}\frac{A_{1}}{A_{0}},\;\;A_{k}=\sum_{i}\tau_{i}^{k}U_{im}U_{in}.
\label{taumntaurelation}
\end{equation}
We will consider separately the probability distribution of these two dynamical
quantities, following Refs.\ \cite{Sch00,Bee99}.

\subsection{Wigner-Smith delay time}
\label{sec21}

There is a close relationship between dynamic scattering problems without
absorption and static problems with absorption \cite{Kly92}. Physically, this
relationship is based on the notion that absorption acts as a ``counter'' for
the delay time of a wave packet \cite{Ram99}. Mathematically, it is based on
the analyticity of the scattering matrix in the upper half of the complex
plane. Absorption with a spatially uniform rate $1/\tau_{\rm a}$ is equivalent
to a shift in frequency by an imaginary amount $\delta\omega={\rm i}/2\tau_{\rm
a}$.\footnote{
To see this, note that absorption is represented by a positive imaginary part
of the dielectric constant $\varepsilon=1+i/\omega\tau_{\rm a}$ (for
$\omega\tau_{\rm a}\gg 1$). Since $\varepsilon$ is multiplied by $\omega^{2}$
in the wave equation, a small imaginary increment
$\omega\rightarrow\omega+i\delta\omega$ is equivalent to absorption with rate
$2\delta\omega$. In the presence of a fluctuating real part of $\varepsilon$,
an imaginary shift in frequency will lead to a spatially fluctuating absorption
rate, but this is statistically equivalent to homogeneous absorption with an
increased scattering rate.}
If we denote the reflection matrix with absorption by $r(\omega,\tau_{\rm a})$,
then $r(\omega,\tau_{\rm a})=r(\omega+{\rm i}/2\tau_{\rm a})$. For weak
absorption we can expand
\begin{equation}
r(\omega+{\rm i}/2\tau_{\rm a})\approx r(\omega)+\frac{\rm i}{2\tau_{\rm
a}}\frac{d}{d\omega}r(\omega)=r(\omega)\left[1-\frac{1}{2\tau_{\rm
a}}Q(\omega)\right].\label{S0Qrelation}
\end{equation}
As before, we have assumed that transmission can be neglected so that $r$ is
unitary and $Q$ is Hermitian. Eq.\ (\ref{S0Qrelation}) implies that the matrix
product $rr^{\dagger}$ for weak absorption is related to the time-delay matrix
$Q$ by a unitary transformation \cite{Bee99},
\begin{equation}
r(\omega,\tau_{\rm a})r^{\dagger}(\omega,\tau_{\rm
a})=r^{\vphantom\dagger}(\omega)\left[1-\frac{1}{\tau_{\rm
a}}Q(\omega)\right]r^{\dagger}(\omega).\label{SQrelation}
\end{equation}

The eigenvalues $R_{1},R_{2},\ldots R_{N}$ of $rr^{\dagger}$ in an absorbing
medium are real numbers between 0 and 1, called the reflection eigenvalues.
Because a unitary transformation leaves the eigenvalues unchanged, one has
$R_{n}=1-\tau_{n}/\tau_{\rm a}$. This relationship between reflection
eigenvalues and Wigner-Smith delay times is useful because the effects of
absorption have received more attention in the literature than dynamic effects.
In particular, the case of a single-mode disordered waveguide with absorption
was solved as early as 1959, in the course of a radio-engineering problem
\cite{Ger59}. The multi-mode case was solved more recently \cite{Bee96,Bru96}.
The distribution is given by the Laguerre ensemble, after a transformation of
variables from $R_{n}$ to $\lambda_{n}=R_{n}(1-R_{n})^{-1}$:
\begin{equation}
P(\{\lambda_{n}\})\propto\prod_{i<j}|\lambda_{i}-\lambda_{j}|^{\beta}
\prod_{k}\exp[-(\alpha\tau_{\rm s}/\tau_{\rm a})(\beta
N+2-\beta)\lambda_{k}].\label{Pwire}
\end{equation}
Here $\tau_{\rm s}$ is the scattering time of the disorder and $\alpha$ is a
numerical coefficient of order unity.\footnote{
The coefficient $\alpha$ depends weakly on $N$ and on the dimensionality of the
scattering: $\alpha=2$ for $N=1$; for $N\rightarrow\infty$ it increases to
$\pi^{2}/4$ or $8/3$ depending on whether the scattering is two or
three-dimensional. The mean free path $l$, that we will encounter later on, is
defined as $l=\alpha'c\tau_{\rm s}$, with $\alpha'=2$ for $N=1$ and
$\alpha'\rightarrow\pi/2$ or $4/3$, respectively, for $N\rightarrow\infty$ in
two or three dimensions. (The wave velocity is denoted by $c$.)
Finally, the diffusion coefficient $D=c^{2}\tau_{\rm
s}/d$ with $d=1$ for $N=1$ and $d\rightarrow 2$ or $3$ for
$N\rightarrow\infty$. The dimensionality that determines these coefficients
is a property of the scattering. It is distinct from the dimensionality of the
geometry. For example, a waveguide geometry (length much greater than width) is
one-dimensional, but it may have $d=3$ (as in the experiments of Ref.\
\protect\cite{Gen99}) or $d=2$ (as in the computer simulations of 
Ref.\ \protect\cite{Sch00}).}
The symmetry index $\beta=1$ in the presence of time-reversal symmetry. (The
case $\beta=2$ of broken time-reversal symmetry is rarely realized in optics.)
The eigenvalue density is given by a sum over Laguerre polynomials, hence the
name ``Laguerre ensemble'' \cite{Meh91}.

The relationship between the reflection eigenvalues for weak absorption and the
Wigner-Smith delay times implies that the $\tau_{n}$'s are distributed
according to Eq.\ (\ref{Pwire}) if one substitutes $\lambda_{n}/\tau_{\rm
a}\rightarrow 1/\tau_{n}$ (since $\lambda_{n}\rightarrow(1-R_{n})^{-1}$ for
weak absorption). In terms of the rates $\mu_{n}=1/\tau_{n}$ one has
\cite{Bee99}
\begin{equation}
P(\{\mu_{n}\})\propto\prod_{i<j}|\mu_{i}-\mu_{j}|^{\beta}
\prod_{k}\exp[-\gamma(\beta N+2-\beta)\mu_{k}].\label{Pmuwire}
\end{equation}
We have abbreviated $\gamma=\alpha\tau_{\rm s}$. For $N=1$ it is a simple
$\beta$-independent exponential distribution \cite{Jay89,Hei90,Com97}, or in
terms of the original variable $\tau$,
\begin{equation}
P(\tau)=2\gamma\tau^{-2}\exp(-2\gamma/\tau).\label{PtauNis1result}
\end{equation}

The slow $\tau^{-2}$ decay gives a logarithmically diverging mean
delay time.  The finite localization length $\xi$ is not sufficient to
constrain the delay time, because of resonant transmission. Resonant
states may penetrate arbitrarily far into the waveguide, and although
these states are rare, they dominate the mean (and higher moments) of
the delay time. The divergence is cut off for any finite length $L$ of
the waveguide. Still, as long as $L\gg\xi$, the resonant states cause
large sample-to-sample fluctuations of the delay times. These large
fluctuations drastically modify the distribution of the single-channel
delay time, as we will discuss next.

\subsection{Single-channel delay time}
\label{sec22}

In view of the relation (\ref{taumntaurelation}), we can compute the
distribution of the single-channel delay time $\phi'$ from that of the
Wigner-Smith delay times, if we also know the distribution of the matrix of
eigenvectors $U$. For a disordered medium it is a good approximation to assume
that $U$ is uniformly distributed in the unitary group, independent of the
$\tau_{n}$'s. The distribution $P(\phi')$ may be calculated analytically in the
regime $N\gg 1$, which is experimentally relevant ($N\simeq 100$ in the
microwave experiments of Ref.\ \cite{Gen99}).

\begin{figure}[tb]
\centerline{
\psfig{figure=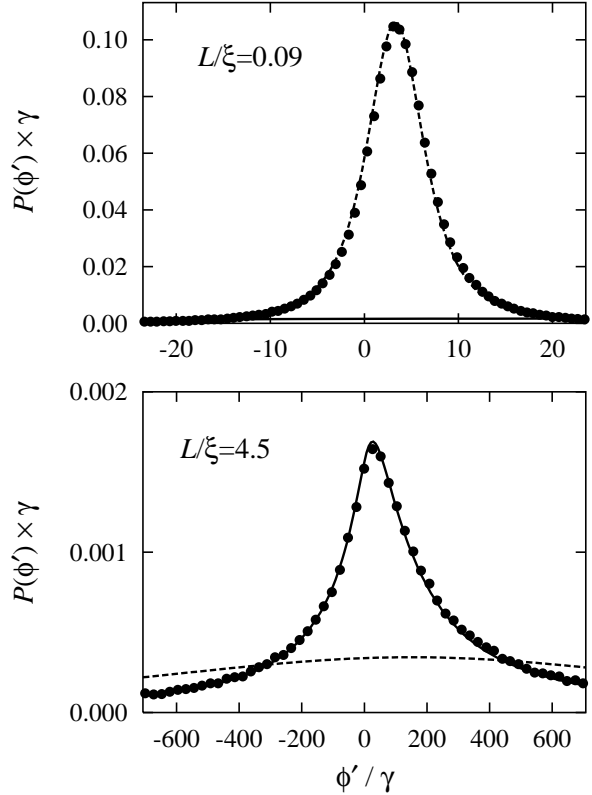,width= 8cm}}
\caption[]{
Distribution of the single-channel delay time $\phi'$ in the diffusive regime
(top panel) and localized regime (bottom panel). The results of numerical
simulations (data points)
are compared to the predictions (\ref{eq:pphi}) (solid curve) and
(\ref{eq:pphidiff}) (dashed). These are results for different incident and
detected modes $n\neq m$. From Ref.\ \protect\cite{Sch00}.
\label{pphidifloc}
}
\end{figure}

In the large-$N$ limit the matrix elements $U_{mn}$ become independent complex
Gaussian random numbers, with zero mean and variance
$\langle|U_{mn}|^{2}\rangle=1/N$. Since Eq.\ (\ref{taumntaurelation}) contains
the elements $U_{im}$ and $U_{in}$, we should distinguish between $n=m$ and
$n\neq m$. Let us discuss first the case $n\neq m$ of different incident and
detected modes. The average over the $U_{in}$'s amounts to doing a set of
Gaussian integrations, with the result \cite{Sch00}
\begin{equation}
P(\phi')=\langle{\textstyle\frac{1}{2}}(B_{2}-B_{1}^{2})
(B_{2}+\phi'^{2}-2B_{1}\phi')^{-3/2}\rangle.
\label{eq:pphi}
\end{equation}
The average $\langle\cdots\rangle$ is over the two spectral moments $B_{1}$ and
$B_{2}$, defined by $B_{k}=\sum_{i}\tau_{i}^{k}|U_{im}|^{2}$. The joint
distribution $P(B_{1},B_{2})$, needed to perform the average, has a rather
complicated form, for which we refer to Ref.\ \cite{Sch00}.

The result (\ref{eq:pphi}) applies to the localized regime $L\gg\xi$. In the
diffusive regime $l\ll L\ll\xi$ one has instead \cite{Gen99,Tig99b}
\begin{equation}
P(\phi')=(Q/2\bar{\phi'})[Q+(\phi'/\bar{\phi'}-1)^{2}]^{-3/2}.
\label{eq:pphidiff}
\end{equation}
The constants are given by $Q\simeq L/l$ and $\bar{\phi'}\simeq L/c$ up to
numerical coefficients of order unity. Comparison of Eqs.\ (\ref{eq:pphi}) and
(\ref{eq:pphidiff}) shows that the two distributions would be identical if
statistical fluctuations in $B_{1}$ and $B_{2}$ could be ignored. However, as a
consequence of the large fluctuations of the Wigner-Smith delay times in the
localized regime, the distribution $P(B_{1},B_{2})$ is very broad and
fluctuations have a substantial effect.

This is illustrated in Fig.\ 4, where we compare $P(\phi')$ in the two regimes.
The data points are obtained from a numerical solution of the wave equation on
a two-dimensional lattice, in a waveguide geometry with $N=50$ propagating
modes. They agree very well with the analytical curves. The distribution
(\ref{eq:pphidiff}) in the diffusive regime decays $\propto|\phi'|^{-3}$, so
that the mean delay time is finite (equal to $\bar{\phi'}$). The distribution
in the localized regime decays more slowly, $\propto|\phi'|^{-2}$. The
resulting logarithmic divergence of the mean delay time is cut off in the 
simulations by the finiteness of the waveguide length.

Notice that, although the most probable value of the single-channel delay time
is positive, the tail of the distribution extends both to positive and negative
values of $\phi'$. This is in contrast to the Wigner-Smith delay time
$\tau_{n}$, which takes on only positive values. The adjective ``delay'' in the
name single-channel delay time should therefore not be taken literally. The
difficulties in identifying the phase derivative with the duration of a
scattering process have been emphasized by B\"{u}ttiker \cite{But90}.

\begin{figure}[tb]
\centerline{
\psfig{figure=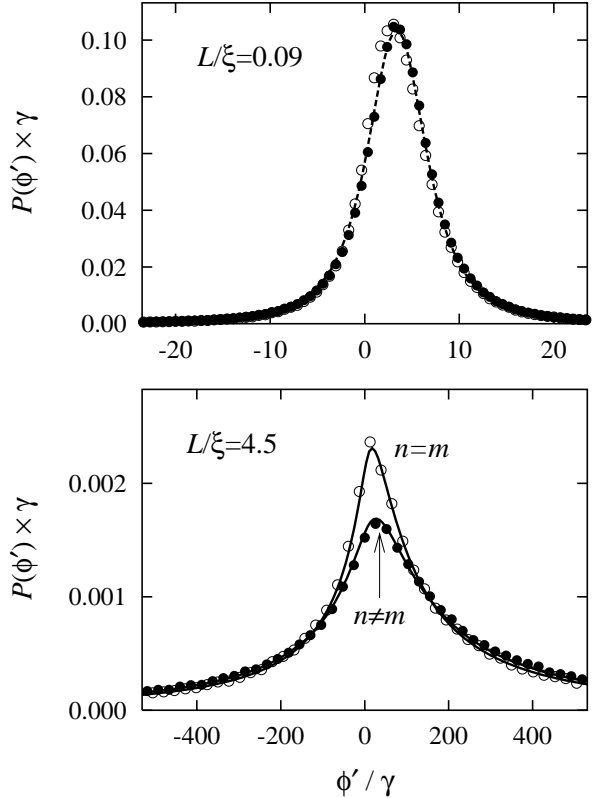,width=
8cm}}
\caption[]{
Same as the previous figure, but now comparing the case $n\neq m$ of different
incident and detected modes (solid circles) with the equal-mode case $n=m$
(open circles). A coherent backscattering effect appears, but only in the
localized regime. From Ref.\ \protect\cite{Sch00}.
\label{pphinism}
}
\end{figure}

We now turn to the case $n=m$ of equal-mode excitation and detection. An
interesting effect of coherent backscattering appears in the localized regime,
as shown in Fig.\ 5. The maximal value of $P(\phi')$ for $n=m$ is larger than
for $n\neq m$ by a factor close to $\sqrt{2}$. (The precise value in the limit
$N\rightarrow\infty$ is $\sqrt{2}\times\frac{4096}{1371\pi}$.) In the diffusive
regime, however, there is no difference in the distributions of the
single-channel delay time for $n=m$ and $n\neq m$.

Coherent backscattering in the original sense is a static scattering property
\cite{Alb85,Wol85}. The distribution $P(I)$ of the reflected intensity differs
if the detected mode is the same as the incident mode or not. The difference
amounts to a rescaling of the distribution by a factor of two,
\begin{equation}
P(I)=\left\{\begin{array}{cc} Ne^{-NI} & \mbox{ if }n\neq m\;,\\
\frac{1}{2}Ne^{-NI/2} & \mbox{ if }n=m\;,
\end{array}
\right. \label{ray}
\end{equation}
so that the mean reflected intensity $\bar{I}$ becomes twice as large near the
angle of incidence. It doesn't matter for this static coherent backscattering
effect whether $L$ is large or small compared to $\xi$. The dynamic coherent
backscattering effect, in contrast, requires localization for its existence,
appearing only if $L>\xi$. This is the dynamical signature of localization
mentioned in the introduction.

\subsection{Transmission}
\label{sec23}

Experiments on the delay-time distribution have so far only been carried out in
transmission, not yet in reflection. The distribution (\ref{eq:pphidiff}) in
the diffusive regime applies both to transmission and to reflection, only the
constants $Q$ and $\bar{\phi'}$ differ \cite{Tig99b}. (In transmission, $Q$ is
of order unity while $\bar{\phi'}\simeq L^{2}/lc$.) Good agreement between
theory and experiment has been obtained both with microwaves \cite{Gen99} and
with light \cite{Lag00}. The microwave data is reproduced in Fig.\ 6.
Absorption can not be neglected in this experiment ($L$ exceeds the absorption
length $l_{\rm a}$ by a factor 2.5), but this can be accounted for simply by a
change in $Q$ and $\bar{\phi'}$. The localization length is larger than $L$ by
a factor of 5, so that the system is well in the diffusive regime. It would be
of interest to extend these experiments into the localized regime, both in
transmission and in reflection. This would require a substantial reduction in
absorption, to ensure that $L<\xi<l_{\rm a}$.

\begin{figure}[tb]
\centerline{
\psfig{figure=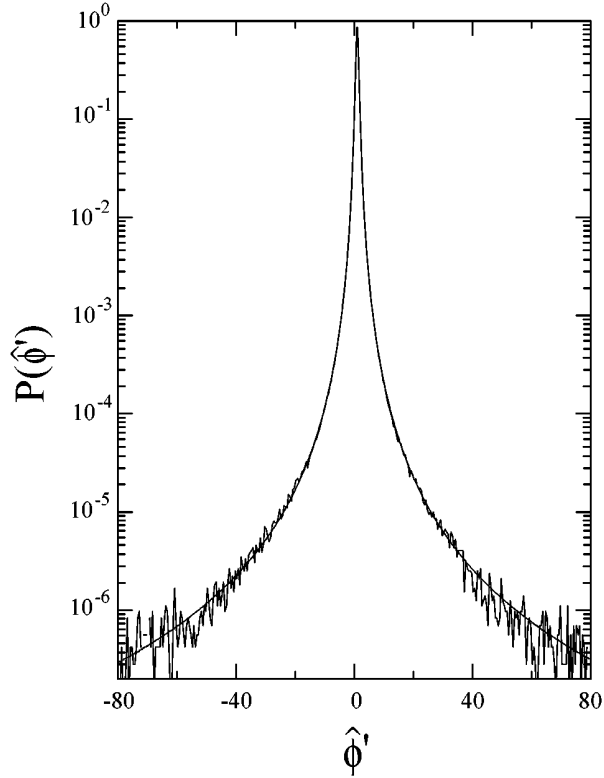,width=
8cm}}
\caption[]{
Distribution of the rescaled single-channel delay time
$\hat{\phi'}=\phi'/\bar{\phi'}$, measured in transmission at a frequency
$\nu\equiv\omega/2\pi=18.1$~GHz on the system described in Fig.\ 3. The smooth
curve through the data is the analytical prediction (\protect\ref{eq:pphidiff})
of diffusion theory (with $Q=0.31$). From Ref.\ \protect\cite{Gen99}.
\label{genack}
}
\end{figure}

Theoretically, much less is known about the delay-time distribution in
transmission than in reflection. While we have a complete theory in reflection,
as described in the previous subsection, in transmission not even the $N=1$
case has been solved completely. Regardless of the value of $N$, one would
expect $P(\phi')$ for $L\gg\xi$ to have the same $1/\phi'^{2}$ tail in
transmission as it has in reflection, since in both cases the same resonances
allow the wave to penetrate deeply into the localized region. For $N=1$ this is
borne out by numerical simulations by Bolton-Heaton et al.\ \cite{Bol99}. These
authors also used a picture of one-dimensional resonant transport through
localized states to study the decay of the weighted delay time $I\phi'$ (with
$I$ the transmitted intensity). They found an algebraic decay for $P(I\phi')$,
just as for $P(\phi')$, but with a different exponent $-4/3$ instead of $-2$.
It is not known how this carries over to $N>1$.

Because of the finite length $L$ of the waveguide, these algebraic tails
are only an intermediate asymptotics. For $N=1$ and exponentially large
times $|\phi'|>\tau_{\rm s}e^{L/l}$ the delay-time distribution has the
more rapid decay \cite{Bol99,Alt88}
\begin{equation}
P(\phi')\propto\exp[-(l/L)\ln^{2}(\phi'/\tau_{\rm s})].\label{Plognormaltail}
\end{equation}
Such a log-normal tail is likely to exist in the multi-mode case as well, but
this has so far only been demonstrated in the diffusive regime $l\ll L\ll\xi$
\cite{Alt91,Muz95,Mir00}. The $1/\phi'^{2}$ intermediate asymptotics does not
appear in that regime.

\section{High-frequency dynamics}

\subsection{Reflection}

The high-frequency limit of the correlator (\ref{Comegardef}) of the reflection
matrices is rather trivial. The two matrices $r^{\dagger}(\omega)$ and
$r(\omega+\delta\omega)$ become uncorrelated for
$\delta\omega\rightarrow\infty$, so that $C$ becomes the product of two
independent random matrices. The distribution of each of these matrices may be
regarded as uniform in the unitary group, and then $C$ is also uniformly
distributed. This is the circular ensemble of random-matrix theory
\cite{Meh91}, so called because the eigenvalues $\exp(i\phi_{n})$ are spread
out uniformly along the unit circle. Their joint distribution is
\begin{equation}
P(\{\phi_{n}\})\propto\prod_{n<m}|e^{i\phi_{n}}-e^{i\phi_{m}}|^{\beta}.
\label{Pcircular}
\end{equation}
This distribution contains no dynamical information.

\subsection{Transmission}

The transmission problem is more interesting at high frequencies. Let us
consider the ensemble-averaged correlator of the transmission matrix elements
\begin{equation}
\langle a_{\omega}(\delta\omega)\rangle=\langle
t^{\ast}_{mn}(\omega)t^{\vphantom\ast}_{mn}(\omega+\delta\omega)\rangle.
\label{Comegatransdef}
\end{equation}
Following Ref.\ \cite{Bee99b}, we proceed as we did in Sec.\ \ref{sec21}, by
mapping the dynamic problem without absorption onto a static problem with
absorption.

We make use of the analyticity of the transmission amplitude
$t_{mn}(\omega+iy)$, at complex frequency $\omega+iy$ with $y>0$, and of the
symmetry relation $t_{mn}(\omega+iy)=t_{mn}^{\ast}(-\omega+iy)$. The product of
transmission amplitudes $t_{mn}(\omega+z)t_{mn}(-\omega+z)$ is an analytic
function of $z$ in the upper half of the complex plane. If we take $z$ real,
equal to $\frac{1}{2}\delta\omega$, we obtain the product
$t_{mn}(\omega+\frac{1}{2}\delta\omega)
t_{mn}^{\ast}(\omega-\frac{1}{2}\delta\omega)$ in Eq.\ (\ref{Comegatransdef})
(the difference with $t_{mn}(\omega+\delta\omega) t_{mn}^{\ast}(\omega)$ being
statistically irrelevant for $\delta\omega\ll\omega$). If we take $z$
imaginary, equal to $i/2\tau_{\rm a}$, we obtain the transmission probability
$T=|t_{mn}(\omega+i/2\tau_{\rm a})|^{2}$ at frequency $\omega$ and absorption
rate $1/\tau_{\rm a}$. We conclude that the ensemble average of $a$ can be
obtained from the ensemble average of $T$ by analytic continuation to imaginary
absorption rate:
\begin{equation}
\langle a_{\omega}(\delta\omega)\rangle=\langle T\rangle\;\;{\rm
for}\;\;1/\tau_{\rm a}\rightarrow -i\delta\omega.\label{CTrelation}
\end{equation}

Higher moments of $a$ are related to higher moments of $T$ by $\langle
a^{p}\rangle=\langle T^{p}\rangle$ for $1/\tau_{\rm a}\rightarrow
-i\delta\omega$. Unfortunately, this is not sufficient to determine the entire
probability distribution $P(a)$, because moments of the form $\langle
a^{p}a^{\ast q}\rangle$ can not be obtained by analytic continuation. This is a
complication of the transmission problem. The reflection problem is simpler,
because the (approximate) unitarity of the reflection matrix $r$ provides
additional information on the distribution of the correlator of the reflection
amplitudes. This explains why in Sec.\ \ref{sec21} we could use the mapping
between the dynamic and absorbing problems to calculate the entire distribution
function of the eigenvalues of $r^{\dagger}(\omega)r(\omega+\delta\omega)$ in
the limit $\delta\omega\rightarrow 0$.

We will apply the mapping first to the single-mode case ($N=1$) and then to the
case $N\gg 1$ of a multi-mode waveguide.

\subsubsection{One mode}
The absorbing problem for $N=1$ was solved by Freilikher, Pustilnik, and
Yurkevich \cite{Fre94}. Applying the mapping (\ref{CTrelation}) to their result
we find\footnote{
The coefficient in front of the factor $L/l$ in the exponent in Eq.\
(\protect\ref{Csingle}) would be $-\frac{1}{3}$ according to the results of
Ref.\ \protect\cite{Fre94}. This would disagree with numerical simulations,
which clearly indicate $|\langle a\rangle|=\exp(-L/l)$ (K. J. H. van Bemmel,
unpublished). The error can be traced back to Eq.\ (39) in Ref.\
\protect\cite{Fre94}.}
\begin{equation}
\langle a_{\omega}(\delta\omega)\rangle=\exp(i\delta\omega L/c-L/l),
\label{Csingle}
\end{equation}
in the regime $c/l\ll\delta\omega\ll (\omega^{2}c/l)^{1/3}$. (The
high-frequency cutoff is due to the breakdown of the random-phase approximation
\cite{Fre97}.) The absolute value $|\langle a\rangle|=\exp(-L/l)$ is
$\delta\omega$-independent in this regime. For $L\ll l$ one has ballistic
motion, hence $\langle a\rangle=\exp(i\delta\omega L/c)$ is simply a phase
factor, with the ballistic time of flight $L/c$. Comparing with Eq.\
(\ref{Csingle}) we see that localization does not change the frequency
dependence of the correlator for large $\delta\omega$, which remains given by
the ballistic time scale, but only introduces a frequency-independent weight
factor.

The implication of this result in the time domain is that $\langle
a_{\omega}(t)\rangle$ has a peak with weight $\exp(-L/l)$ at the ballistic time
$t=L/c$. Such a ballistic peak is expected for the propagation of classical
particles through a random medium, but it is surprising to find that it applies
to wave dynamics as well.

\subsubsection{Many modes}
Something similar happens for $N\gg 1$. The transmission probability in an
absorbing multi-mode waveguide was calculated by Brouwer \cite{Bro98},
\begin{equation}
\langle T\rangle=\frac{l}{N\xi_{\rm a}\sinh(L/\xi_{\rm
a})}\exp\left(-\frac{L}{2Nl}\right),\label{Tmulti}
\end{equation}
for absorption lengths $\xi_{\rm a}=\sqrt{D\tau_{\rm a}}$ in the range
$l\ll\xi_{\rm a}\ll\xi$. The length $L$ of the wave\-guide should be $\gg l$,
but the relative magnitude of $L$ and $\xi$ is arbitrary. Substitution of
$1/\tau_{\rm a}$ by $-i\delta\omega$ gives the correlator
\begin{equation}
\langle
a_{\omega}(\delta\omega)\rangle=\frac{l\sqrt{-i\tau_{D}\delta\omega}}
{NL\sinh\sqrt{-i\tau_{D}\delta\omega}}\exp\left(-\frac{L}{2Nl}\right), 
\label{Cmulti}
\end{equation}
where $\tau_{D}=L^{2}/D$ is the diffusion time. The range of validity of Eq.\
(\ref{Cmulti}) is $L/\xi\ll\sqrt{\tau_{D}\delta\omega}\ll L/l$, or equivalently
$D/\xi^{2}\ll\delta\omega\ll c/l$. In the diffusive regime, for $L\ll\xi$, the
correlator (\ref{Cmulti}) reduces to the known result \cite{Ber94} from
perturbation theory.

For ${\rm max}\,(D/L^{2},D/\xi^{2})\ll\delta\omega\ll c/l$ the decay of the
absolute value of the correlator is a stretched exponential,
\begin{equation}
|\langle
a_{\omega}(\delta\omega)\rangle|=\frac{2l}{NL}
\sqrt{\tau_{D}\delta\omega}\exp\left(-\sqrt{{\textstyle\frac{1}{2}}
\tau_{D}\delta\omega}-\frac{L}{2Nl}\right).\label{Cstretched}
\end{equation}
In the localized regime, when $\xi$ becomes smaller than $L$, the onset of this
tail is pushed to higher frequencies, but it retains its functional form. The
weight of the tail is reduced by a factor $\exp(-L/2Nl)$ in the presence of
time-reversal symmetry. (There is no reduction factor if time-reversal symmetry
is broken \cite{Bee99b}.)

\begin{figure}
\centerline{
\psfig{figure=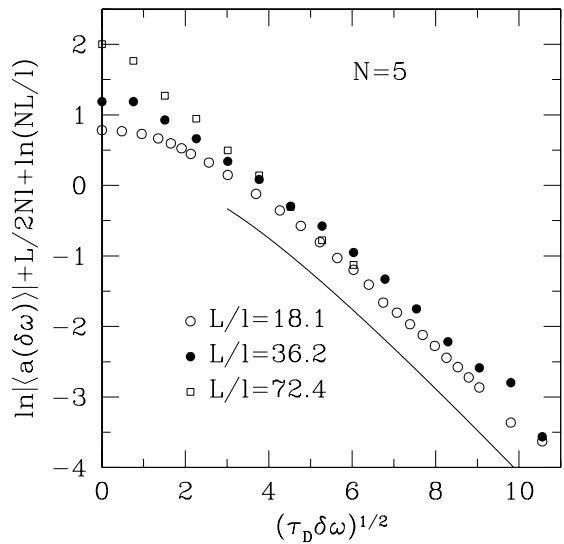,width= 8cm}}
\medskip
\caption[]{
Frequency dependence of the logarithm of the absolute value of the correlator
$\langle a_{\omega}(\delta\omega)\rangle$. The data points follow from a
numerical simulation for $N=5$, the solid curve is the analytical
high-frequency result (\protect\ref{Cstretched}) for $N\gg 1$. The decay of the
correlator is given by the diffusive time constant $\tau_{D}=L^{2}/D$ even if
the length $L$ of the wave\-guide is greater than the localization length
$\xi=6\,l$. The offset of about 0.6 between the numerical and analytical
results is probably a finite-$N$ effect. From Ref.\ \protect\cite{Bee99b}.
\label{Cmultiplot}}
\end{figure}

In Fig.\ 7 we compare the results of numerical simulations in a two-dimensional
waveguide geometry with the analytical high-frequency prediction. We see that
the correlators for different values of $L/\xi$ converge for large
$\delta\omega$ to a curve that lies somewhat above the theoretical prediction.
The offset is probably due to the fact that $N$ is not $\gg 1$ in the
simulation. Regardless of this offset, the simulation confirms both analytical
predictions: The stretched exponential decay
$\propto\exp(-\sqrt{\tau_{D}\delta\omega/2})$ and the exponential suppression
factor $\exp(-L/2\xi)$. We emphasize that the time constant $\tau_{D}=L^{2}/D$
of the high-frequency decay is the diffusion time for {\em the entire length\/}
$L$ of the waveguide --- even though the localization length $\xi$ is up to a
factor of 12 smaller than $L$.

We can summarize these findings \cite{Bee99b} for the single-mode and
multi-mode waveguides by
the statement that the correlator of the transmission amplitudes {\em
factorises\,} in the high-frequency regime: $\langle
a_{\omega}(\delta\omega)\rangle\rightarrow f_{1}(\delta\omega)f_{2}(\xi)$. The
frequency dependence of $f_{1}$ depends on the diffusive time through the
wave\-guide, even if it is longer than the localization length. Localization
has no effect on $f_{1}$, but only on $f_{2}$.

\section{Propagation of a pulse}

If the incident wave is a short pulse, then the separation into low- and
high-frequency dynamics is less natural. Ideally one would like to know the
entire time dependence of the correlator $a_{\omega}(t)$ introduced in Eq.\
(\ref{Comegatdef}). A complete solution exists \cite{Tit00} for the
ensemble-averaged correlator in the case of reflection,
\begin{eqnarray}
\langle
a_{\omega}(t)\rangle&=&\int_{-\infty}^{\infty}
\frac{d\delta\omega}{2\pi}\,e^{-i\delta\omega t}\langle r^{\ast}_{mn}(\omega)r^{\vphantom\ast}_{mn}(\omega+\delta\omega)
\rangle\nonumber\\
&=&\frac{1+\delta_{mn}}{N(N+1)}\int_{-\infty}^{\infty}
\frac{d\delta\omega}{2\pi}\,e^{-i\delta\omega t}\langle {\rm Tr}\,C_{\omega}(\delta\omega)\rangle.\label{ameandef}
\end{eqnarray}
The second equality follows from the representation
\begin{equation}
r(\omega\pm\delta\omega/2)=U^{\rm T}e^{\pm i\Phi/2}U,\label{rUPhiU}
\end{equation}
with $\Phi={\rm diag}(\phi_{1},\phi_{2},\ldots\phi_{N})$ a diagonal matrix and
$U$ uniformly distributed in the unitary group. The factor $1+\delta_{mn}$ is
due to coherent backscattering. It is convenient to work with the normalized
power spectrum,
\begin{equation}
{\cal P}_{\omega}(t)=\frac{N+1}{1+\delta_{mn}}\langle
a_{\omega}(t)\rangle=\frac{1}{N}\int_{-\infty}^{\infty}
\frac{d\delta\omega}{2\pi}\,e^{-i\delta\omega t}\langle {\rm Tr}\,C_{\omega}(\delta\omega)\rangle,\label{calPdef}
\end{equation}
normalized such that $\int_{0}^{\infty} dt\,{\cal P}_{\omega}(t)=1$.

Since $e^{i\phi_{n}}$ is an eigenvalue of the unitary matrix $C$, one can write
\begin{equation}
\langle {\rm
Tr}\,C_{\omega}(\delta\omega)\rangle=\int_{0}^{2\pi}d\phi\,\rho(\phi)e^{i\phi},
\label{Crhoaverage}
\end{equation}
where $\rho(\phi)=\langle\sum_{n=1}^{N}\delta(\phi-\phi_{n})\rangle$ is the
phase-shift density. This density can be obtained from the corresponding
density $\rho(R)$ of reflection eigenvalues $R_{n}$ (eigenvalues of
$rr^{\dagger}$) in an absorbing medium, by analytic continuation to imaginary
absorption rate: $i/\tau_a\rightarrow\delta\omega$,
$R_{n}\rightarrow\exp(i\phi_n)$. The densities are related by
\begin{equation}
\rho(\phi)=\frac{N}{2\pi}+\frac{1}{\pi}{\rm
Re}\,\sum_{n=1}^{\infty}e^{-in\phi}\int_{0}^{1}R^{n}\rho(R)\,dR,
\label{continuation}
\end{equation}
as one can verify by equating moments. This is a quick and easy way to solve
the problem, since the probability distribution of the reflection eigenvalues
is known \cite{Bee96,Bru96}: it is given by the Laguerre ensemble
(\ref{Pwire}). The density $\rho(R)$ can be obtained from that as a series of
Laguerre polynomials, using methods from random-matrix theory \cite{Meh91}.
Eq.\ (\ref{continuation}) then directly gives the density $\rho(\phi)$.

One might wonder whether one could generalize Eq.\ (\ref{continuation}) to
reconstruct the entire distribution function $P(\{\phi_{n}\})$ from the
Laguerre ensemble of the $R_{n}$'s. The answer is no, unless $\delta\omega$ is
infinitesimally small (as in Sec.\ \ref{sec21}). The reason that the method of
analytic continuation can not be used to obtain correlations between the
$\phi_{n}$'s is that averages of negative powers of $\exp(i\phi_n)$ are not
analytic in the reflection eigenvalues. For example, for the two-point
correlation function one would need to know the average
$\langle\exp(i\phi_n-i\phi_m)\rangle\rightarrow\langle R_{n}R_{m}^{-1}\rangle$
that diverges in the absorbing problem. It {\em is\/} possible to compute
$P(\{\phi_{n}\})$ for any $\delta\omega$ --- but that requires a different
approach, for which we refer to Ref.\ \cite{Tit00}.

The calculation of the power spectrum from Eqs.\
(\ref{calPdef})---(\ref{continuation}) is easiest in the absence of
time-reversal symmetry, because $\rho(R)$ then has a particularly simple form.
One obtains the power spectrum \cite{Tit00}
\begin{eqnarray}
{\cal P}_{\omega}(t)&=&-\frac{1}{N}\frac{d}{dt}F(t/2N\gamma),\label{agen}\\
F(t)&=&\frac{1}{t+1}\sum_{n=0}^{N-1}\left(\frac{t-1}{t+1}\right)^{n}
P_{n}\left(\frac{t^{2}+1}{t^{2}-1}\right),\label{Fdef}
\end{eqnarray}
where $P_{n}$ is a Legendre polynomial. (Recall that $\gamma=\alpha\tau_{\rm
s}$, cf.\ Sec.\ \ref{sec21}.) In the single-mode case Eq.\ (\ref{agen})
simplifies to \cite{Whi87}
\begin{equation}
{\cal P}_{\omega}(t)=2\gamma(t+2\gamma)^{-2}.\label{alocalized}
\end{equation}
It decays as $t^{-2}$. For $N\rightarrow\infty$ Eq.\ (\ref{agen}) simplifies to
\begin{equation}
{\cal P}_{\omega}(t)=t^{-1}\exp(-t/\gamma)I_{1}(t/\gamma),\label{adiffusive}
\end{equation}
where $I_{1}$ is a modified Bessel function. The power spectrum now decays as
$t^{-3/2}$. For any finite $N$ we find a crossover from ${\cal
P}=\sqrt{\gamma/2\pi}\,t^{-3/2}$ for $\tau_{\rm s}\ll t\ll N^{2}\tau_{\rm s}$
to ${\cal P}=2N\gamma t^{-2}$ for $t\gg N^2\tau_{\rm s}$. This is illustrated
in Fig.\ 8.

\begin{figure}
\centerline{
\psfig{figure=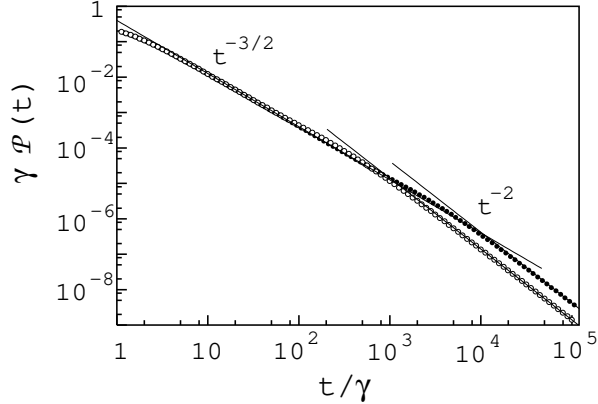,width= 8cm}}
\medskip
\caption[]{
Time dependence of the power spectrum of a reflected pulse in the absence of
time-reversal symmetry, calculated from Eq.\ (\protect\ref{agen}) for $N=7$
(open circles) and $N=21$ (filled circles). The intermediate-time asymptote
$\propto t^{-3/2}$ and the large-time asymptote $\propto t^{-2}$ are shown as
straight lines in this double-logarithmic plot. The prefactor is
$N$-independent for intermediate times but $\propto N$ for large times (notice
the relative offset of the large-time asymptotes). Courtesy of M. Titov.
\label{powerlaws}}
\end{figure}

In the presence of time-reversal symmetry the exact expression for ${\cal
P}_{\omega}(t)$ is more cumbersome but the asymptotics carries over with minor
modifications. In particular, the large-$N$ limit (\ref{adiffusive}) with its
$t^{-3/2}$ decay remains the same, while the $t^{-2}$ decay changes only in the
prefactor: ${\cal P}=(N+1)\gamma t^{-2}$ for $t\gg N^2\tau_{\rm s}$.

The quadratic tail of the time-dependent power spectrum of a pulse 
reflected from an infinitely long waveguide is the same as the quadratic
tail of the delay-time distribution that we have encountered in
Sec.\ \ref{sec2}. It is natural to assume that the power spectrum for
transmission through a localized waveguide of finite length has the same
quadratic decay, with a cross-over to a log-normal tail for exponentially
large times, cf.\ Sec.\ \ref{sec23}.

\section{Conclusion}

We now have a rather complete picture of the dynamics of a wave reflected by a
disordered waveguide. The dynamical information is contained in the phase
factors $e^{i\phi_{n}}$ that are the eigenvalues of the product of reflection
matrices $r^{\dagger}(\omega)r(\omega+\delta\omega)$. Three regimes can be
distinguished, depending on the magnitude of the length scale
$l_{\delta\omega}=\sqrt{D/\delta\omega}$ associated with the frequency
difference $\delta\omega$:
\begin{itemize}
\item Ballistic regime, $l_{\delta\omega}<l$. This is the high-frequency
regime. The statistics of the $\phi_{n}$'s is given by the circular ensemble,
Eq.\ (\ref{Pcircular}).
\item Localized regime, $l_{\delta\omega}>\xi$. This is the low-frequency
regime. The $\phi_{n}$'s are now distributed according to the Laguerre ensemble
(\ref{Pmuwire}).
\item Diffusive regime, $l<l_{\delta\omega}<\xi$. The distribution of the
$\phi_{n}$'s does not belong to any of the known ensembles of random-matrix
theory \cite{Tit00}.
\end{itemize}

The emphasis in this review has been on the localized regime. The dynamics is
then dominated by resonances that allow the wave to penetrate deep into the
waveguide. Such resonances correspond to large delay times
$\tau_{n}=\lim_{\delta\omega\rightarrow 0}\phi_{n}/\delta\omega$. The
distribution of the largest delay time $\tau_{\rm max}$ follows from the
distribution of the smallest eigenvalue in the Laguerre ensemble \cite{Ede91}.
For $\beta=1$ it is given by
\begin{equation}
P(\tau_{\rm max})=\gamma N(N+1)\tau_{\rm max}^{-2} \exp(-\gamma
N(N+1)/\tau_{\rm max}).\label{Ptaumax}
\end{equation}
It has a long-time tail $\propto 1/\tau_{\rm max}^{2}$, so that the mean
delay time diverges (in the limit of an infinitely long waveguide). A
subtle and unexpected consequence of the resonances is the appearance
of a dynamic coherent backscattering effect in the distribution of the
single-mode delay times. Unlike the conventional coherent backscattering
effect in the static intensity, the dynamic effect requires localization
for its existence. The recent progress in time-resolved measurements of
light scattering from random media, reported at this meeting \cite{Lag00},
should enable observation of this effect.

Extension of the theory to two- and three-dimensional localization remains a
challenging problem for future research. We believe that the dynamic coherent
backscattering effect will persist in higher dimensions, provided the
localization length remains large compared to the mean free path. Several
methods have been proposed to distinguish absorption from localization in the
static intensity \cite{Cha00,Tig00}. The effect reviewed here could provide
this information from a different, dynamic, perspective.

\section*{Acknowledgments}
It is a pleasure to acknowledge the fruitful collaboration on this topic with
K. J. H. van Bemmel, P. W. Brouwer, H. Schomerus, and M. Titov. This research
was supported by the ``Ne\-der\-land\-se or\-ga\-ni\-sa\-tie voor
We\-ten\-schap\-pe\-lijk On\-der\-zoek'' (NWO) and by the ``Stich\-ting voor
Fun\-da\-men\-teel On\-der\-zoek der Ma\-te\-rie'' (FOM).



\begin{references}

\bibitem{She90} {\em Scattering and Localization of Classical Waves in Random
Media}, edited by P. Sheng (World Scientific, Singapore, 1990).
\bibitem{Tig99} B. A. van Tiggelen, in {\em Diffuse Waves in Complex Media},
edited by J.-P. Fouque, NATO Science Series C531 (Kluwer, Dordrecht, 1999).
\bibitem{Whi87} B. White, P. Sheng, Z. Q. Zhang, and G. Papanicolaou, Phys.\
Rev.\ Lett.\ {\bf 59}, 1918 (1987).
\bibitem{Lag00} A. Lagendijk, J. G\'{o}mez Rivas, A. Imhof, F. J. P.
Schuurmans, and R. Sprik, in this volume.
\bibitem{Nat99} F. Scheffold, R. Lenke, R. Tweer, and G. Maret, Nature {\bf
398}, 206 (1999); D. S. Wiersma, J. G\'{o}mez Rivas, P. Bartolini, A.
Lagendijk, and R. Righini, Nature {\bf 398}, 207 (1999).
\bibitem{Sch00} H. Schomerus, K. J. H. van Bemmel, and C. W. J. Beenakker, {\tt
cond-mat/0004049}; {\tt cond-mat/0009014}.
\bibitem{Efe97} K. Efetov, {\em Supersymmetry in Disorder and Chaos\/}
(Cambridge University, Cambridge, 1997).
\bibitem{Bee97} C. W. J. Beenakker, Rev.\ Mod.\ Phys.\ {\bf 69}, 731 (1997).
\bibitem{Asc91} M. Asch, W. Kohler, G. Papanicolaou, M. Postel, and B. White,
SIAM Review {\bf 33}, 519 (1991).
\bibitem{Seb99} P. Sebbah, O. Legrand, and A. Z. Genack, Phys.\ Rev.\ E {\bf
59}, 2406 (1999).
\bibitem{Gen99} A. Z. Genack, P. Sebbah, M. Stoytchev, and B. A. van Tiggelen,
Phys.\ Rev.\ Lett.\ {\bf 82}, 715 (1999).
\bibitem{Wig55} E. P. Wigner, Phys.\ Rev.\ {\bf 98}, 145 (1955).
\bibitem{Smi60} F. T. Smith, Phys.\ Rev.\ {\bf 118}, 349 (1960).
\bibitem{Bee99} C. W. J. Beenakker and P. W. Brouwer, {\tt cond-mat/9908325}.
\bibitem{Kly92} V. I. Klyatskin and A. I. Saichev, Usp.\ Fiz.\ Nauk {\bf 162},
161 (1992) [Sov.\ Phys.\ Usp.\ {\bf 35}, 231 (1992)].
\bibitem{Ram99} S. A. Ramakrishna and N. Kumar, Phys.\ Rev.\ B {\bf 61}, 3163
(2000).
\bibitem{Ger59} M. E. Gertsenshtein and V. B. Vasil'ev, Teor.\ Veroyatn.\
Primen.\ {\bf 4}, 424 (1959); {\bf 5}, 3(E) (1960) [Theor.\ Probab.\ Appl.\
{\bf 4}, 391 (1959); {\bf 5}, 340(E) (1960)].
\bibitem{Bee96} C. W. J. Beenakker, J. C. J. Paasschens, and P. W. Brouwer,
Phys.\ Rev.\ Lett.\ {\bf 76}, 1368 (1996).
\bibitem{Bru96} N. A. Bruce and J. T. Chalker, J. Phys.\ A {\bf 29}, 3761
(1996); {\bf 29}, 6681(E) (1996).
\bibitem{Meh91} M. L. Mehta, {\em Random Matrices} (Academic, New York, 1991).
\bibitem{Jay89} A. M. Jayannavar, G. V. Vijayagovindan, and N. Kumar, Z. Phys.\
B {\bf 75}, 77 (1989).
\bibitem{Hei90} J. Heinrichs, J. Phys.\ Condens.\ Matter {\bf 2}, 1559 (1990).
\bibitem{Com97} A. Comtet and C. Texier, J. Phys. A {\bf 30}, 8017 (1997).
\bibitem{Tig99b} B. A. van Tiggelen, P. Sebbah, M. Stoytchev, and A. Z. Genack,
Phys.\ Rev.\ E {\bf 59}, 7166 (1999).
\bibitem{But90} M. B\"{u}ttiker, in {\em Electronic Properties of Multilayers
and Low-Dimensional Semiconductor Structures}, edited by J. M. Chamberlain, L.
Eaves, and J. C. Portal, NATO ASI Series B231 (Plenum, New York, 1990).
\bibitem{Alb85} M. P. van Albada and A. Lagendijk, Phys.\ Rev.\ Lett.\ {\bf
55}, 2692 (1985).
\bibitem{Wol85} P.-E. Wolf and G. Maret, Phys.\ Rev.\ Lett.\ {\bf 55}, 2696
(1985).
\bibitem{Bol99} C. J. Bolton-Heaton, C. J. Lambert, V. I. Fal'ko, V. Prigodin,
and A. J. Epstein, Phys.\ Rev.\ B {\bf 60}, 10569 (1999).
\bibitem{Alt88} B. L. Altshuler and V. N. Prigodin, Pis'ma Zh.\ Eksp.\ Fiz.\
{\bf 47}, 36 (1988) [JETP Lett.\ {\bf 47}, 43 (1988)].
\bibitem{Alt91} B. L. Altshuler, V. E. Kravtsov, and I. V. Lerner, in {\em
Mesoscopic Phenomena in Solids}, edited by B. L. Altshuler, P. A. Lee, and R.
A. Webb (North-Holland, Amsterdam, 1991).
\bibitem{Muz95} B. A. Muzykantskii and D. E. Khmelnitskii, Phys.\ Rev.\ B {\bf
51}, 5480 (1995).
\bibitem{Mir00} A. D. Mirlin, Phys.\ Rep.\ {\bf 326}, 259 (2000).
\bibitem{Bee99b} C. W. J. Beenakker, K. J. H. van Bemmel, and P. W. Brouwer,
Phys.\ Rev.\ E {\bf 60}, R6313 (1999).
\bibitem{Fre94} V. Freilikher, M. Pustilnik, and I. Yurkevich, Phys.\ Rev.\ B
{\bf 50}, 6017 (1994).
\bibitem{Fre97} V. Freilikher and M. Pustilnik, Phys.\ Rev.\ B {\bf 55}, R653
(1997).
\bibitem{Bro98} P. W. Brouwer, Phys.\ Rev.\ B {\bf 57}, 10526 (1998).
\bibitem{Ber94} R. Berkovits and S. Feng, Phys.\ Rep.\ {\bf 238}, 135 (1994).
\bibitem{Tit00} M. Titov and C. W. J. Beenakker, {\tt cond-mat/0005042}.
\bibitem{Ede91} A. Edelman, Lin.\ Alg.\ Appl.\ {\bf 159}, 55 (1991).
\bibitem{Cha00} A. A. Chabanov, M. Stoytchev, and A. Z. Genack, Nature {\bf
404}, 850 (2000).
\bibitem{Tig00} B. A. van Tiggelen, A. Lagendijk, and D. S. Wiersma, Phys.\
Rev.\ Lett.\ {\bf 84}, 4333 (2000).

\end{references}
\end{document}